\documentclass[
 reprint,
superscriptaddress,
 amsmath,amssymb,
prb,
floatfix,
]{revtex4-2}

\usepackage{graphicx}
\usepackage{dcolumn}
\usepackage{bm}
\usepackage{xcolor}
\usepackage{float}
\usepackage[colorlinks,urlcolor=blue,citecolor=blue,linkcolor=blue]{hyperref}

\begin{document}

\preprint{APS/123-QED}

\title{Quantum geometric tensor and wavepacket dynamics in two-dimensional non-Hermitian systems}

\author{Y.-M. Robin Hu}
\affiliation{%
Department of Quantum Science and Technology, Research School of Physics, The Australian National University, Canberra, ACT 2601 Australia
}%
\author{Elena A. Ostrovskaya}%
\affiliation{%
Department of Quantum Science and Technology, Research School of Physics, The Australian National University, Canberra, ACT 2601 Australia
}%
\author{Eliezer Estrecho}

\affiliation{%
Department of Quantum Science and Technology, Research School of Physics, The Australian National University, Canberra, ACT 2601 Australia
}%


\begin{abstract}
The quantum geometric tensor (QGT) characterizes the local geometry of quantum states, and its components directly account for the dynamical effects observed, e.g., in condensed matter systems. In this work, we address the problem of extending the QGT formalism to non-Hermitian systems with gain and loss. In particular, we investigate a wave-packet dynamics in two-band non-Hermitian systems to elucidate how non-Hermiticity affects the definition of QGT.
We employ first-order perturbation theory to account for non-adiabatic corrections due to interband mixing. Our results suggest that two different generalizations of the QGT, one defined using only the right eigenstates and the other one using both the left and right eigenstates, both play a significant role in wave-packet dynamics. We then determine the accuracy of the perturbative approach by simulating a wave-packet dynamics in a well studied physical non-Hermitian system -- exciton polaritons in a semiconductor microcavity. Our work aids deeper understanding of quantum geometry and dynamical behaviour in non-Hermitian systems.
\end{abstract}

\maketitle


\emph{Introduction.}---
The complex-valued quantum geometric tensor (QGT) was first proposed by Provost and Vallee in 1980 to quantify the distance (in parameter space) between two quantum states \cite{provost1980}. Its real (symmetric) part is the quantum metric tensor (QMT) describing the overlap between two quantum states and its imaginary (anti-symmetric) part is the Berry curvature describing the topology of the system's energy bands. In periodic systems, e.g. solid-state lattices, the integral of the Berry curvature over a Brillouine zone is a topological invariant known as the Chern number \cite{provost1980,berry1984,klitzing1980,laughlin1981,thouless1982,avron1983,niu1985,xiao2010,bleu2018}. The Berry curvature acts on the wave-packet centre of mass (e.g., an electron) as an effective magnetic field in momentum space, and results in an anomalous velocity transverse to the direction of the applied force \cite{sundaram1999,culcer2005,gao2014,gao2015WP,bleu2018WP,gianfrate2020,chang1995,chang2008}. The QMT appears in the wave-packet equation of motions as perturbative corrections to the Berry curvature and describes the field induced positional shift \cite{gao2014,gao2015WP} and non-adiabatic effects \cite{bleu2018WP}. It also plays an important role in flat-band superconductivity \cite{peotta2015,tian2023,chen2024} and can also be assopciated with fidelity susceptibility that has been used to detect criticality, which plays an important role in quantum information theory \cite{jozsa1994,gu2010,wang2015,wei2018,khan2024}.

Generalizing the QGT to systems effectively described by non-Hermitian Hamiltonians ~\cite{hatano1996,bender1998,bender1999,brody2014,el-ganainy2018,ghatak2019,bergoltz2021,ozdemir2019} is of fundamental importance for understanding the dynamical effects in systems with gain and loss. Non-Hermitian systems have attracted a great amount of interests because they can feature novel topological invariants \cite{gong2018}, topological phases \cite{gong2018,kawabata2019}, and topological edge states \cite{kunst2018,leykam2017,borgnia2020,zhang2020,zhang2022,lee2019,pernet2022,kokhanchik2023,manna2023}. Non-Hermitian effects have also been experimentally studied in a variety of physical systems including cold atoms \cite{graefe2008,liang2022}, photonics \cite{weidemann2020,longhi2022,guria2024,zhong2021,zhong2023}, mechanical systems \cite{ghatak2020}, electric circuits \cite{hofmann2020}, magnons \cite{mcclarty2019,yu2024}, spin liquid \cite{yang2022}, and exciton polaritons~\cite{gao2018,su2021}.

The challenge in generalizing the QGT lies in the fact that non-Hermitian Hamiltonians have distinct left and right eigenstates~\cite{el-ganainy2018,ghatak2019,bergoltz2021,ozdemir2019}. Some works therefore define the QGT using only the right eigenstates, which we denote the right-right (RR) formalism \cite{solnyshkov2021,liao2021,cuerda2024,cuerda2024ob,matsumoto2020,alon2024}, while others use both the left and right eigenstates, which we denote the left-right (LR) formalism \cite{brody2013,zhang2019,zhu2021,tzeng2021,fan2020,ye2023}. Previous studies proposed that it is the RR version for the Berry curvature that describes the anomalous velocity in the dynamics of a wave-packet centre of mass \cite{xu2017,silberstein2020,wang2022}. However, there are also non-Hermitian corrections arising from the Berry connections defined in both the RR and the LR formalism \cite{xu2017,silberstein2020,wang2022}. At the same time, while Refs.~\cite{solnyshkov2021,alon2024} suggests that the RR QMT plays a significant role in the wave-packet dynamics,  Ref.~\cite{ye2023} states that the RR QMT is ill-defined due to the non-orthogonality of the right eigenstates. So far, it remains unclear which version of the QMT correctly represents the physical (measurable) quantities of non-Hermitian systems and how well the QMT describes the dynamical effects in non-Hermitian systems.

In this work, we address this ambiguity in the QGT definition by investigating the wave-packet dynamics of a two-band non-Hermitian system using the first-order perturbative theory.

We derive a semi-classical equation of the center-of-mass motion of a wave-packet, which explicitly shows the contributions of both the LR and RR components of the QGT. We then compared the analytical results with a numerical simulation using a model describing an experimentally accessible non-Hermitian system. The model describes two-dimensional hybrid light-matter particles in a solid state -- microcavity exciton polaritons~\cite{carusotto2013}. Both real and imaginary parts of the eigenenergies~\cite{su2021} as well as QGT components ~\cite{gianfrate2020,hu2024} can be measured in this system, in principle allowing for experimental verification of theoretical predictions.

\emph{Theory.---}
For an $n$-th level quantum state $|u_n\rangle$ that is smoothly dependent on momentum, the QGT can be defined as~\cite{provost1980}:
\begin{equation}
        Q_{n,ij}=\langle\partial_{k_i} u_n|\partial_{k_j} u_n\rangle-\langle \partial_{k_i} u_n|u_n\rangle\langle u_n|\partial_{k_j} u_n\rangle
\end{equation}
where $k_i, k_j$ denote the indices in momentum space. The QGT can also be written in terms of inter-band Berry connections $\mathbf{A}_{nm}=\langle u_n|i\partial_\mathbf{k}u_m\rangle$, in the form of $Q_{n,ij}=\sum_{m\neq n}A_{nm,i}A_{mn,j}$ \cite{bleu2018WP,piechon2016}. The QMT, $g_{n,ij}=\operatorname{Re}[Q_{n,ij}]$ and the Berry curvature, $\Omega_{n,ij}=-2\operatorname{Im}[Q_{n,ij}]$ correspond to the real and imaginary parts of the QGT, respectively.

In non-Hermitian systems, where the Hamiltonian $H\neq H^\dagger$, the left and right eigenstates defined as:
\begin{equation}
        H|u_n^R\rangle=E_n|u_n^R\rangle,  \qquad
        \langle u_n^L|H=\langle u_n^L|E_n
\end{equation}
are not equal but are biorthogonal, satisfying $\langle u_n^L|u_m^R\rangle=0$ for $m\neq n$ \cite{bergoltz2021,ghatak2019}. This distinction between the left and right eigenstates leads to different ways of defining the QGT. The QGT can be defined using only the right eigenstates, like in Refs. \cite{solnyshkov2021,liao2021,cuerda2024,cuerda2024ob,matsumoto2020}. In this formalism, the inner product is defined as $\langle \cdot\rangle=\langle u_n^R|\cdot|u_n^R\rangle$ and the QGT and its components can be written as:
\begin{equation}
    \begin{split}
        Q_{n,ij}^{RR}=&\langle\partial_{k_i} u_n^R|\partial_{k_j} u_n^R\rangle-\langle \partial_{k_i} u_n^R|u_n^R\rangle\langle u_n^R|\partial_{k_j} u_n^R\rangle,\\
        g_{n,ij}^{RR}=&\operatorname{Re}[Q_{n,ij}^{RR}],  \qquad
        \Omega_{n,ij}^{RR}=-2\operatorname{Im}[Q_{n,ij}^{RR}].
    \end{split}
\end{equation}
In what follows, we will denote the components of the QGT defined this way as the right-right (RR) generalization. Similarly to the Hermitian case, the RR QMT and the RR Berry curvature correspond to the real and imaginary parts of the RR QGT, and both quantities are real-valued.

Alternatively, the QGT can also be defined using both the left and the right eigenstates, like in Refs. \cite{zhang2019,zhu2021,tzeng2021,fan2020,ye2023}, where the inner product is defined using biorthogonality, $\langle \cdot\rangle=\langle u_n^L|\cdot|u_n^R\rangle$. The corresponding QGT and its components are
\begin{equation}
\label{eq:LRdef}
    \begin{split}
        Q_{n,ij}^{LR}=&\langle\partial_{k_i} u_n^L|\partial_{k_j} u_n^R\rangle-\langle \partial_{k_i} u_n^L|u_n^R\rangle\langle u_n^L|\partial_{k_j} u_n^R\rangle,\\
        g_{n,ij}^{LR}=&\frac{1}{2}(Q_{n,ij}^{LR}+Q_{n,ji}^{LR}),\\
        \Omega_{n,ij}^{LR}=&i(Q_{n,ij}^{LR}-Q_{n,ji}^{LR}).
    \end{split}
\end{equation}
We refer to this formalism as the left-right (LR) generalization. Unlike the Hermitian case, the LR QMT and the LR Berry curvature can be complex-valued, and correspond to the symmetric and the anti-symmetric parts of the LR QGT. Note that this is not the only way to define a LR generalization of the QMT. Ref.~\cite{ye2023} proposed two other possible generalizations such as defining the LR QMT as the real part of the LR QGT or the real and symmetric part of the LR QGT. Here, we use Eqs.~(\ref{eq:LRdef}) to maintain the symmetric properties in the Hermitian case and our results support this definition, as will be shown below. Note that, in both formalisms, we define the eigenstates such that $\langle u_n^L|u_n^R\rangle=\langle u_n^R|u_n^R\rangle=1$ to ensure that both $Q_n^{RR}$ and $Q_n^{LR}$ are gauge-invariant.

The relevance of the components of the QGT to the dynamics has been investigated in both Hermitian and non-Hermitian systems. In the Hermitian case, the semi-classical equations of motion of a narrow wave-packet in a single band can be written as ~\cite{gao2014,gao2015WP,bleu2018WP}:
\begin{equation}\label{eq: hermitian eom g}
    \begin{split}
    \Dot{\mathbf{k}}_c&=\mathbf{F}/\hbar\\
    \Dot{\mathbf{r}}_c&=\partial_\mathbf{k}\epsilon_0-\mathbf{F}\times\Big(\mathbf{\Omega}_0+\nabla_\mathbf{k}\times\frac{2g_0\cdot\mathbf{F}}{\epsilon_1-\epsilon_0}\Big),
    \end{split}
\end{equation}
where $\mathbf{k}_c$ and $\mathbf{r}_c$ are the center-of-mass (COM) momentum and position, respectively, and $\mathbf{F}$ is an external force. The first term, the gradient of the energy $\epsilon_0$ of the band, is the group velocity. The second term is a field-induced correction due to the (vector) Berry curvature $\mathbf{\Omega}_0$, resulting in the anomalous Hall drift~\cite{gianfrate2020,bleu2018WP,bleu2018,leblanc2021,piechon2016} perpendicular to the field $\mathbf{F}$. The last term is the result of first-order perturbation theory accounting for non-adiabatic corrections due to a neighboring band with energy $\epsilon_1$. This term is second order in $\mathbf{F}$ and is proportional to the QMT $g_0$.

In the non-Hermitian case, only the first-order field-induced corrections were previously worked out~\cite{xu2017,silberstein2020,wang2022} with the COM position written as:
\begin{equation}
\label{eq: rdot0}
       \Dot{\mathbf{r}}_c=\partial_{\mathbf{k}}\operatorname{Re}\Big[\epsilon_0+\mathbf{F}\cdot(\mathbf{A}_{00}^{RR}-\mathbf{A}_{00}^{LR})\Big]-\mathbf{F}\times\mathbf{\Omega}_0^{RR}.
\end{equation}
Here, the real part of the complex-valued eigen-energy $\epsilon_0$ determines the group velocity and the RR Berry curvature determines the anomalous Hall drift. The non-Hermitian contribution is due to the real part of the difference of the RR and LR intra-band Berry connections of the band $A_{00}^{RR}$, $A_{00}^{LR}$ defined as
\begin{equation}
    \begin{split}
       \mathbf{A}_{nn}^{RR}=\langle u_n^R|i\partial_\mathbf{k}u_n^R\rangle, \qquad
        \mathbf{A}_{nn}^{LR}=\langle u_n^L|i\partial_\mathbf{k}u_n^R\rangle.
    \end{split}
\end{equation}
Note that $\mathbf{A}_{nn}^{LR}$ is generally complex-valued, but only the real part is relevant in Eq.~(\ref{eq: rdot0}). The Berry connection $\mathbf{A}_{nn}^{RR}$ is real-valued when the right eigenstates are normalized as $\langle u_n^R|u_n^R\rangle=1$ \cite{xu2017,silberstein2020,hu2024}.

The non-Hermitian anomalous Berry connection \cite{xu2017,wang2022} $(\mathbf{A}_{nn}^{RR}-\mathbf{A}_{nn}^{LR})$ results in a field-induced correction to the energy\cite{silberstein2020}. In the Hermitian limit, this term in the equation of motion vanishes and $\mathbf{\Omega}^{RR}\rightarrow \mathbf{\Omega}$, reducing Eq.~(\ref{eq: rdot0}) to the Hermitian case~\cite{sundaram1999}, Eq.~\ref{eq: hermitian eom g}, without the adiabatic correction.

In this work, we derive the second-order field-induced corrections to find out how the non-Hermitian QGT components affect the wave-packet dynamics by including interband mixing.

\emph{Results.}---
We consider a perturbation to a two-dimensional (2D) Hamiltonian in the form of
\begin{equation}
    \tilde H=H-\mathbf{F}\cdot\mathbf{r},
\end{equation}
where $\mathbf{F}$ represents an external force and $\mathbf{r}$ is the position operator.
We then consider a Gaussian wave packet in the state
\begin{equation}
    |W\rangle=\int w(\mathbf{k},t)e^{i\mathbf{k}\cdot\mathbf{r}}|\tilde u_0^R(\mathbf{k})\rangle d^2\mathbf{k},
\end{equation}

where $|\tilde u_0^R\rangle$ is a right eigenstate of the Hamiltonian $\tilde H$ that includes the external force.

In this work, we focus on two-band systems with gain and loss and denote $|\tilde u_0^R\rangle$ to be the eigenstate with the larger growth (or the smaller decay) rate. The wave packet will eventually evolve into the eigenstate with the largest $\operatorname{Im}E_n$ due to the loss or gain \cite{silberstein2020,hu2022}. Therefore, in order to employ a single-band approximation, we need to ensure that the wave packet stays in the eigenstate that has the larger $\operatorname{Im}E_n$, i.e. larger gain (or smaller loss).

We then apply the perturbation theory ~\cite{sternheim1972} by expanding the eigenstate to the first order correction in $\mathbf{F}$ as,
\begin{equation}
    |\tilde u_0^R(\mathbf{k})\rangle=|u_0^R(\mathbf{k})\rangle-\frac{\mathbf{F}\cdot \mathbf{A}_{10}^{LR}}{\epsilon_0-\epsilon_1}|u_1^R(\mathbf{k})\rangle
\end{equation}
where $|u_1^R(\mathbf{k})\rangle$ is the other eigenstate (with larger loss), $\mathbf{A}_{10}^{LR}=\langle u_1^L|i\partial_\mathbf{k}u_0^R\rangle$ is the LR inter-band Berry connection, and $\epsilon_{0,1}$ are the complex-valued energy eigenvalues of the unperturbed Hamiltonian $H$. We emphasize that the inter-band Berry connection in the perturbative correction has to be defined in the ``left-right'', instead of the conventional ``right-right'', formalism due to biorthogonality in non-Hermitian systems  \cite{sternheim1972}.

Assuming that the wave packet is narrowly centered at its center of mass in momentum space $|w(\mathbf{k},t)|^2\approx\delta(\mathbf{k}-\mathbf{k}_c)$, we then follow the formalism presented in Refs.~\cite{xu2017,wang2022} and derive the equation of motion of the wave packet COM in position space up to the first-order perturbative correction (see SI for details) as:
\begin{equation}\label{eq: corrected eom}
    \begin{split}       \hbar\Dot{\mathbf{r}}_c=&\partial_{\mathbf{k}}\operatorname{Re}\Big[\epsilon_0+\mathbf{F}\cdot(\tilde{\mathbf{A}}_{00}^{RR}-\tilde{\mathbf{A}}_{00}^{LR})\Big]-\mathbf{F}\times\tilde{\mathbf{\Omega}}_0^{RR}
        \end{split}
\end{equation}
where,
\begin{align}
\tilde{\mathbf{A}}_{00}^{RR}=&\mathbf{A}_{00}^{RR}+\operatorname{Re}\bigg[\frac{2Q_{0}^{RR}\cdot\mathbf{F}}{\epsilon_1-\epsilon_0}\bigg]\notag\\
&-\operatorname{Im}\bigg[\partial_\mathbf{k}\Big(\frac{\mathbf{F}\cdot(\mathbf{A}_{00}^{RR}-\mathbf{A}_{00}^{LR})}{\epsilon_1-\epsilon_0}\Big)\bigg] \label{eq:RRcorr}\\
\tilde{\mathbf{A}}_{00}^{LR}=&\mathbf{A}_{00}^{LR}+ \frac{2g^{LR}\cdot\mathbf{F}}{\epsilon_1-\epsilon_0} \label{eq:LRcorr} \\
\tilde{\mathbf{\Omega}}_0^{RR}=&\mathbf{\Omega}_0^{RR}+\nabla_\mathbf{k}\times\operatorname{Re}\bigg[\frac{2Q_{0}^{RR}\cdot\mathbf{F}}{\epsilon_1-\epsilon_0}\bigg] \label{eq:BCcorr}.
\end{align}

These expressions form the main result of our work: Eq.~(\ref{eq: corrected eom}) is the non-Hermitian generalization of Eq.~(\ref{eq: hermitian eom g}) and Eqs.~(\ref{eq:RRcorr}, \ref{eq:LRcorr}, \ref{eq:BCcorr}), explicitly showing the first-order perturbation corrections to Eq.~(\ref{eq: rdot0}). In the following, we will show that the second-order (in $\mathbf{F}$) corrections have contributions from the QMT, Berry curvature, and anomalous non-Hermitian Berry connection.

Let us first examine the correction to the Berry curvature $\tilde{\mathbf{\Omega}}_0^{RR}$ by comparing Eq.~(\ref{eq: corrected eom}) to Eq.~(\ref{eq: hermitian eom g}). The second-order field-induced correction clearly transforms as $g_0/(\epsilon_1-\epsilon_0)\rightarrow \operatorname{Re}[Q_0^{RR}/(\epsilon_1-\epsilon_0)]$. Note, however, that this is not simply $g_0\rightarrow g_0^{RR}$ due to the complex denominator. In fact, there is an additional non-Hermitian correction which comes from the RR Berry curvature itself, as shown below:
\begin{equation}
    \operatorname{Re}\Big[\frac{2Q_0^{RR}\cdot\mathbf{F}}{\Delta\epsilon}\Big]=\operatorname{Re}\left[\frac{1}{\Delta\epsilon}\right]2g_{0}^{RR}\cdot\mathbf{F}
   -   \operatorname{Im}\left[\frac{1}{\Delta\epsilon}\right]\mathbf{F}\times\mathbf{\Omega}_0^{RR}
\end{equation}
This new term depends on the imaginary part of the complex energy gap, which naturally vanishes in the Hermitian limit or along the imaginary Fermi arc, where $\operatorname{Im}\Delta\epsilon=0$~\cite{su2021,hu2022}.

The second-order field-induced corrections to the anomalous non-Hermitian Berry connection have two components, i.e. $\operatorname{Re}[\mathbf{F}\cdot(\tilde{\mathbf{A}}_{00}^{RR}-\tilde{\mathbf{A}}_{00}^{LR})]=I_0+I_1 +  I_2$, where $I_0=\mathbf{F}\cdot(\mathbf{A}_{00}^{RR}-\mathbf{A}_{00}^{LR})]$ is the first-order term. The first correction is proportional to the difference of the RR and LR QMTs and has the form
\begin{equation}
    I_1 = 2\mathbf{F}\cdot\operatorname{Re}\left[\frac{g_0^{RR}-g_0^{LR}}{\Delta\epsilon}\right]\cdot\mathbf{F},
\end{equation}
while the second term is proportional to the anomalous Berry connection and has the form
\begin{equation}
    I_2 = -\mathbf{F}\cdot\operatorname{Im}\bigg[\partial_\mathbf{k}\Big(\frac{\mathbf{F}\cdot(\mathbf{A}_{00}^{RR}-\mathbf{A}_{00}^{LR})}{\Delta\epsilon}\Big)\bigg].
\end{equation}
Note that both the real and imaginary parts of the LR QMT and Berry connections contribute to the corrections due to the complex denominator (see SI for the explicit forms of $I_1$ and $I_2$). This novel non-Hermitian effect shows that the LR QMT should indeed be defined as a complex-valued quantity \cite{zhang2019,zhu2021,ye2023}. This is in contrast to the corrections to the Berry curvature, Eq.~(\ref{eq:BCcorr}), which originate from the real-valued RR QMT and Berry curvature.

To examine the corrections closely, we integrate Eq.~\ref{eq: corrected eom} to get (see SI for the derivation):
 \begin{align}
         \mathbf{r}_c=&\bigg[-\partial_\mathbf{k}\varphi+\tilde{\mathbf{A}}_{00}^{RR}\bigg]_{\mathbf{k}=\mathbf{k}_c},\label{eq:rc}\\
\varphi=&-\frac{1}{\hbar}\int_{t_0}^{t}\operatorname{Re}\bigg[ \epsilon_0-\mathbf{F}\cdot\tilde{\mathbf{A}}_{00}^{LR}\bigg]_{\mathbf{k}=\mathbf{k}_c}dt'.\label{eq:phi}
     \end{align}
The positional shift~\cite{gao2014} of the COM, given by Eq.~(\ref{eq:rc}), is due to the gradient of the phase $\varphi$ and the corrected RR Berry connection. The latter, which is given by Eq.~(\ref{eq:RRcorr}), has two field-induced corrections. The first depends on the RR QGT, $Q_0^{RR}$, which is the same term that appears in field-induced correction to the Berry curvature $\tilde{\mathbf{\Omega}}_0^{RR}$ in Eq.~(\ref{eq:BCcorr}). The second correction depends on the anomalous non-Hermitian Berry connection, $\mathbf{A}_{00}^{RR}-\mathbf{A}_{00}^{LR}$, which also appears in the field-induced correction to the energy in Eq.~(\ref{eq: rdot0}). There is also a field-induced correction to the geometric (Berry) phase~\cite{berry1984,xu2017,wang2022}, Eq.~(\ref{eq:phi}), which is proportional to the LR QMT, $g_0^{LR}$.

In the Hermitian limit, the eigenenergies become real-valued, $\mathbf{A}_{00}^{LR}=\mathbf{A}_{00}^{RR}$, and $Q_0^{RR}=Q_0^{LR}$ becoming Hermitian as well. Therefore, the correction to the LR and RR Berry connections both take the forms of $\mathbf{A}_{00}\rightarrow \mathbf{A}_{00}+2(g\cdot\mathbf{F})/(\Delta\epsilon)$, and we recover the previous results given by Eq.~(\ref{eq: hermitian eom g})~\cite{gao2014, gao2015WP,bleu2018WP}.

\begin{figure}[t]
    \centering
    \includegraphics[width=0.5\textwidth]{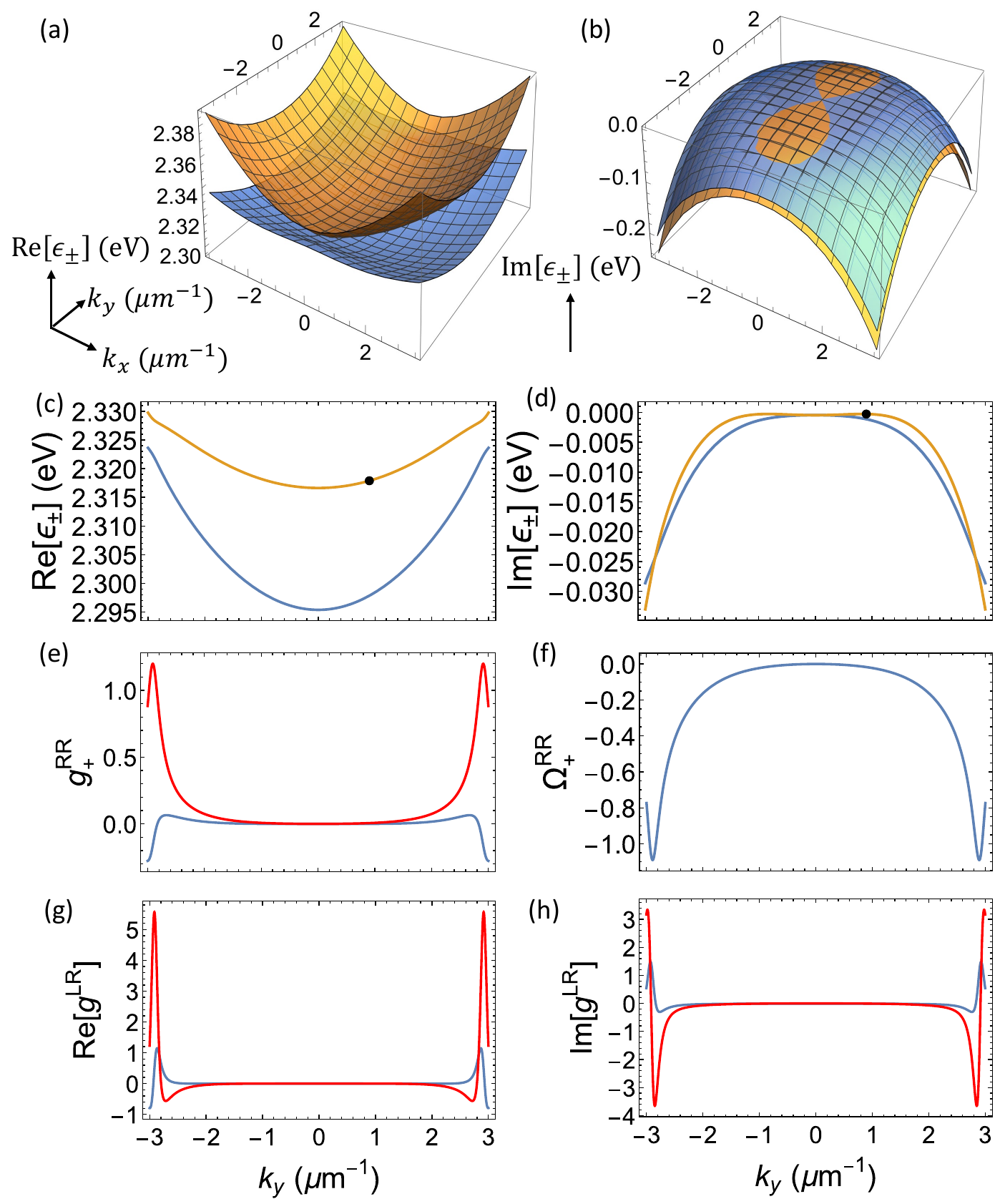}
    \caption{The (a) real and (b) imaginary parts of the eigenenergies of the exciton polaritons in a gapped phase with $\Delta_z=0.007$ eV. (c, d) The cross-sections of the real and imaginary parts of the eigenenergies along the $k_y$-axis, respectively. The orange line and surface represent the upper band, the blue line and surface represent the lower band, and the black dots represent the initial momentum of the wave-packet center-of-mass. (e, f) The RR quantum metric and Berry curvature. (g, h) The real and imaginary parts of the LR quantum metric. The blue and red lines in (e, g, h) show the $xy$-components and the $yy$-components respectively, since these are the only components that appears in the dynamics when the force is in the $y$-direction.
    }
    \label{fig: e_qgt}
\end{figure}

To demonstrate the relevance of the QGT components to the description of dynamical effects in a two-band non-Hermitian system, we simulate the wave-packet dynamics in an exciton-polariton system formed by strong coupling of an optically active, 2D-confined exciton to a photon confined in an optical microcavity. It is a 2D (two-dimensional) system featuring rich non-Hermitian effects due to inherent loss and gain~\cite{gao2015}.  The non-Hermiticity of the system arises from the decay of polaritons, mainly due to photons leaking out through the imperfect mirrors of the microcavity. Due to a pseudospin (polarisation) degree of freedom, this system can also feature non-trivial bands with
exceptional points in momentum space~\cite{su2021,hu2022,hu2024}. In principle, exciton-polaritons can provide an experimental platform for verifying the results presented here since components of both the QGT \cite{bleu2018,gianfrate2020} and its non-Hermitian generalization~\cite{hu2024} can be measured in this system.

We employ split-step methods \cite{weideman1986,taha2005} to verify that the first order perturbation theory developed here improves the accuracy of the semi-classical equation of motion.
We consider a two-band non-Hermitian effective Hamiltonian for exciton polaritons in a form of a complex-valued effective magnetic field $\overrightarrow{\mathbf{G}}=[G_x,G_y,G_z]$ acting on the exciton-polariton pseudospins described by the $2\times2$ Pauli matrices $\overrightarrow{\sigma}=[\sigma_x,\sigma_y,\sigma_z]$. The Hamiltonian has the form~\cite{su2021,hu2022}
\begin{equation}\label{eq: exciton polariton H}
    \begin{split}
        H(\mathbf{k})&=H_0(\mathbf{k})+\overrightarrow{\mathbf{G}}(\mathbf{k})\cdot\overrightarrow{\sigma}\\
        H_0(\mathbf{k})&=E_0+\frac{\hbar^2 k^2}{2m}-i\gamma_0-i\gamma_2 k^2-i\gamma_4 k^4\\
        \overrightarrow{\mathbf{G}}(\mathbf{k})&=[\tilde\alpha+\tilde\beta(k_x^2-k_y^2),2k_x k_y\tilde\beta,\Delta_z],
    \end{split}
\end{equation}
where $m$ is the effective exciton-polariton mass and $\gamma$ represent the mean exciton-polariton linewidth expanded to the fourth order in $k$, which arises from the experimentally observed non-parabolic dependence of the linewidth on momentum \cite{su2021}. Here, $\tilde\alpha=\alpha-ia$ and $\tilde\beta=\beta-ib$, where $\alpha$ denotes the splitting of the linearly-polarized modes due to the cavity anisotropy, $\beta$ describes the TE-TM splitting as a result of the photonic spin-orbit coupling, while $a,b$ represent the splitting of the exciton-polariton linewidths due to the polarization-dependent photonic losses and $\Delta_z$ is the Zeeman splitting which arises from an external magnetic field and splits the circularly-polarized modes \cite{su2021,tercas2014}. In the simulations \cite{fig2data}, we set the values of the parameters to be $E_0-i\gamma_0=(2.31-4.5\times10^{-4}i)$ eV, $\hbar^2/(2m)=2.3\times 10^{-3}$ 
eV$\mu$m$^2$, $\gamma_2=0$ eV$\mu$m$^2$, $\gamma_4=3.75\times10^{-4}$ eV$\mu$m$^{4}$, $\tilde\alpha=(8\times 10^{-3}-10^{-5}i)$ eV, $\tilde\beta=(10^{-3}-7.5\times 10^{-4}i)$ eV$\mu$m$^{2}$, which corresponds to the experimentally relevant values. Figure~\ref{fig: e_qgt}(a-d) shows the complex eigenenergies as a function of the momentum $\mathbf{k}$ (see also SI). 

We initialized the wave-packet at $\mathbf{k}=(0,0.9)\mu$m$^{-1}$ in the upper eigenstate which has the larger $\operatorname{Im}E$ (or less loss) [see the dot in Figs.~\ref{fig: e_qgt}(c,d)]. The two bands are separated at this point in momentum space and $\partial_{k_x}\operatorname{Re}\epsilon_0=0$. Experimentally, the wave-packet can be prepared using resonant optical excitation, where the exciting laser wavelength is adjusted to match the energy and the excitation angle is tuned to match the momentum of the target state~\cite{gianfrate2020}.

\begin{figure}[t]
    \centering
    \includegraphics[width=0.5\textwidth]{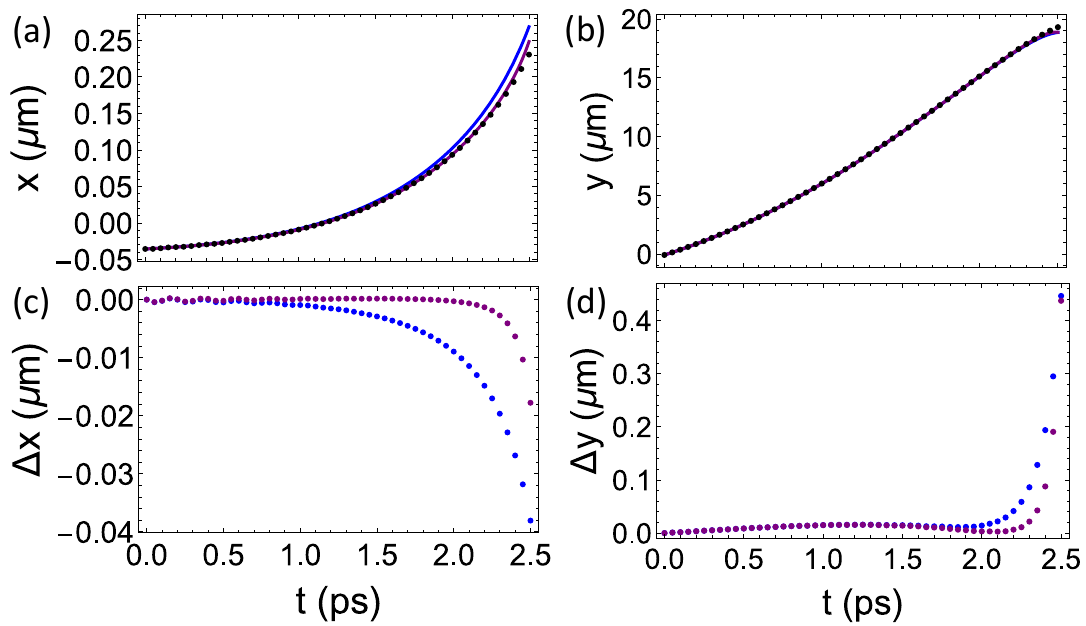}
    \caption{(a, d) The real-space trajectory of exciton-polariton wave-packets initially placed in the upper eigenstate in 3 different gapped phases with $\Delta_z=0.007$ eV. Shown are the simulated results (black dots), the solution of semi-classical equation of motion derived previously Eq.~(\ref{eq: rdot0}) (blue lines) and the solution of the semi-classical equation of motion derived here Eq.~(\ref{eq: corrected eom}) (purple lines). (c, d) The difference between the analytical results and the numerical simulation with the same color scheme as in (a, b).} 
    \label{fig: wp_qgt}
\end{figure}

An external force $\mathbf{F}=(0,5\times10^{-4})$ eV/$\mu$m is applied accelerating the wave-packet along the $+k_y$-axis towards smaller (real) energy spacing. In an optical microcavity, this force naturally arises from the small variation or tilt in the cavity length introduced by the sample fabrication process, which translates to a linear energy gradient experienced by polaritons \cite{steger2015}. As the bands get closer, the mixing between the two eigenstates will be more profound and can result in oscillation due to the zitterbewegung effect \cite{sedov2018,leblanc2021}. However, since the wave packet evolves into the less dissipative eigenstate, the inter-band mixing will gradually disappear and so will the oscillation. We also ensure that the wave packet does not cross the imaginary Fermi arc~\cite{su2021}, where the imaginary parts of the eigenenergies cross, otherwise the single-band approximation will fail~\cite{hu2022} as the loss rates of the eigenstates will switch. The numerical results are presented as dots in Figs.~\ref{fig: wp_qgt}.

In this simple case, Eq.~(\ref{eq: corrected eom}) becomes:
\begin{align}
            v_x=&\frac{1}{\hbar}\partial_{k_x}\operatorname{Re}\Big[F_y\Big(\tilde{A}_{00,y}^{RR}-\tilde{A}_{00,y}^{LR}\Big)\Big]-\frac{F_y}{\hbar}\tilde{\Omega}_{0,xy}^{RR}\\
            v_y=&\frac{1}{\hbar}\partial_{k_y}\operatorname{Re}\Big[\epsilon_0+F_y\Big(\tilde{A}_{00,y}^{RR}-\tilde{A}_{00,y}^{LR}\Big)\Big].
    \end{align}
The explicit forms of $\tilde{A}_{00,y}^{RR}-\tilde{A}_{00,y}^{LR}$ and $\tilde{\Omega}_{0,xy}^{RR}$ are presented in the SI.

The solution of Eq.~(\ref{eq: corrected eom}) gives a better fit to the numerical results than the solution of Eq.~(\ref{eq: rdot0}). This is apparent from the discrepancy between the numerical and analytical solutions plotted in  Figs.~\ref{fig: wp_qgt} (c,d). Although the nonadiabatic correction is rather small in our example, a larger force or a smaller energy gap will result in a much larger correction, meaning that the previously derived equation of motion without perturbative corrections  Eq.~(\ref{eq: rdot0}) will no longer be adequately describing the dynamics. In this case, however, we might also need to extend our treatment beyond the first-order perturbation theory and consider higher order corrections. On the other hand, if the wave packet propagates for a longer time, it would cross the imaginary Fermi arc, and the semi-classical equations of motion will no longer be valid~\cite{hu2022}.

\emph{Conclusion.}---
We derive the first-order perturbative corrections to the semi-classical equation of motion describing the centre-of-mass motion of a wave-packet of a non-Hermitian two-band system. Our results show that both the left-right (LR) and right-right (RR) quantum geometric tensor (QGT) play a role in the dynamics. In particular, the RR QGT describes the non-Hermitian generalization of the field-induced positional shift, while the LR QGT describes the field-induced correction to the Berry phase. Interestingly, our results also suggest novel second-order anomalous Hall drift correction arising from the RR Berry curvature, and the positional-shift correction arising from the difference between the RR and LR QMT and the anomalous Berry connection. More importantly, our theoretical results show that the imaginary part of the LR QMT and Berry connections play a direct role in the second-order field-induced corrections, unlike the first-order case. Our results settle the dispute on whether the LR or the RR QGT should be used to describe the dynamics of non-Hermitian systems, and instead suggest that both LR and RR QGTs represent physical quantities and play a significant role in the dynamics of non-Hermitian systems.

We further confirmed the accuracy of our results by simulating the center-of-mass dynamics of a non-Hermitian exciton-polariton system with a gapped spectrum, which can be realized in microcavity exciton-polaritons exposed to a strong out-of-plane magnetic field~\cite{klembt2018,polimeno2021tuning} or hosted in  liquid-crystal cavities~\cite{lempicka2022electrically}.

Our work opens up many possibilities for further investigation in the QGT in non-Hermitian systems. In particular, our results should be relevant for the physics of second-harmonic generation \cite{bhalla2022}, non-linear response theory \cite{das2023,wang2023}, quantum information \cite{jozsa1994,gu2010} and flat-band superconductivity \cite{peotta2015,tian2023,chen2024}, where the quantum metric tensor plays a significant role. 

\hfill

We acknowledge support from the Australian Research Council (ARC) through the Centre of Excellence Grant CE170100039 and the Discovery Early Career Researcher Award DE220100712, and Australian Government Research Training Program (RTP) Scholarship.

\end{document}


\preprint{APS/123-QED}

\title{Supplementary Material - Quantum geometric tensor and wavepacket dynamics in two-dimensional non-Hermitian systems}

\author{Y.-M. Robin Hu}
\affiliation{%
Department of Quantum Science and Technology, Research School of Physics, The Australian National University, Canberra, ACT 2601 Australia
}%
\author{Elena A. Ostrovskaya}%
\affiliation{%
Department of Quantum Science and Technology, Research School of Physics, The Australian National University, Canberra, ACT 2601 Australia
}%
\author{Eliezer Estrecho}

\affiliation{%
Department of Quantum Science and Technology, Research School of Physics, The Australian National University, Canberra, ACT 2601 Australia
}%

\maketitle

\section{Identities of the Quantum Geometric Tensors}
In this section, we provide identities related to the left-right and the right-right quantum geometric tensor to aide the derivations in the next section. The components of the left-right quantum geometric tensor (LR QGT) is defined as
\begin{equation*}\tag{S1}
    Q_{n,ij}^{LR}=\langle\partial_{k_i} u_n^L|\partial_{k_j} u_n^R\rangle-\langle \partial_{k_i} u_n^L|u_n^R\rangle\langle u_n^L|\partial_{k_j} u_n^R\rangle,
\end{equation*}
where the left and right eigenstates satisfy the normalization $\langle u_n^L|u_n^R\rangle=\langle u_n^R|u_n^R\rangle=1$. We can also define the intra-band and inter-band Berry connections as $\mathbf{A}_{nm}^{\alpha\beta}=\langle u_n^\alpha|i\partial_\mathbf{k}u_m^\beta\rangle$ with $\alpha,\beta=L,R$. 

The LR QGT can be written in terms of the inter-band Berry connection \cite{zhang2019} by inserting $\sum_m|u_m^R\rangle\langle u_m^L|=\mathbf{I}$ to get
\begin{equation*}\tag{S2}
    \begin{split}
        Q_{n,ij}^{LR}=&\sum_m\langle\partial_{k_i} u_n^L|u_m^R\rangle \langle u_m^L|\partial_{k_j} u_n^R\rangle-\langle \partial_{k_i} u_n^L|u_n^R\rangle\langle u_n^L|\partial_{k_j} u_n^R\rangle\\
        &=\sum_{m\neq n}\langle\partial_{k_i} u_n^L|u_m^R\rangle \langle u_m^L|\partial_{k_j} u_n^R\rangle\\
        &=\sum_{m\neq n}\langle-i\partial_{k_i} u_n^L|u_m^R\rangle \langle u_m^L|i\partial_{k_j} u_n^R\rangle\\
         &=\sum_{m\neq n}\langle u_n^L|i\partial_{k_i} u_m^R\rangle \langle u_m^L|i\partial_{k_j} u_n^R\rangle\\
         &=\sum_{m\neq n}(A_{nm}^{LR})_i(A_{mn}^{LR})_j.
         \end{split}
\end{equation*}
We use the identity $\langle \partial_{k_i} u_m^L|u_n^R\rangle=-\langle u_m^L|\partial_{k_i} u_n^R\rangle$ on the third line, based on:
\begin{equation*}\tag{S3}
    \begin{split}
        \partial_{k_i}\langle u_n^L|u_m^R\rangle&=0\\
        \langle \partial_{k_i} u_n^L|u_m^R\rangle + \langle u_n^L|\partial_{k_i} u_m^R\rangle&=0\\
        -\langle \partial_{k_i} u_n^L|u_m^R\rangle &= \langle u_n^L|\partial_{k_i} u_m^R\rangle.
    \end{split}
\end{equation*}
Note that this also implies that $(\mathbf{A}_{nm}^{LR})^*=\mathbf{A}_{mn}^{RL}$. In a two-band system, the LR QGT simplifies to
$$
 Q_{n,ij}^{LR} = A_{nm,i}^{LR}A_{mn,j}^{LR}.
$$

As in the main text, the LR quantum metric tensor (QMT) corresponds to the symmetric part of the LR QGT, therefore, its components can be written as
\begin{equation*}\tag{S4}
    g_{n,ij}^{LR}=\sum_{m\neq n}\frac{1}{2}(A_{nm}^{LR})_i(A_{mn}^{LR})_j +\frac{1}{2}(A_{nm}^{LR})_j(A_{mn}^{LR})_i.
\end{equation*}
Similarly, the components of the right-rght quantum geometric tensor (RR QGT) is defined as
\begin{equation*}\tag{S5}
    Q_{n,ij}^{RR}=\langle\partial_{k_i} u_n^R|\partial_{k_j} u_n^R\rangle-\langle \partial_{k_i} u_n^R|u_n^R\rangle\langle u_n^R|\partial_{k_j} u_n^R\rangle.
\end{equation*}
Where we can use the completeness relation $\sum_m |u_m^L\rangle\langle u_m^R|=\mathbf{I}$ and rewrite the RR QGT as
\begin{equation*}\tag{S6}
\begin{split}
    Q_{n,ij}^{RR}=&\sum_m \langle\partial_{k_i} u_n^R|u_m^L\rangle\langle u_m^R|\partial_{k_j} u_n^R\rangle-\langle \partial_{k_i} u_n^R|u_m^L\rangle\langle u_m^R|u_n^R\rangle\langle u_n^R|\partial_{k_j} u_n^R\rangle\\
    =&\langle\partial_{k_i} u_n^R|u_n^L\rangle\langle u_n^R|\partial_{k_j} u_n^R\rangle-\langle \partial_{k_i} u_n^R|u_n^L\rangle\langle u_n^R|\partial_{k_j} u_n^R\rangle+\sum_{m\neq n} \langle\partial_{k_i} u_n^R|u_m^L\rangle\langle u_m^R|\partial_{k_j} u_n^R\rangle-\langle \partial_{k_i} u_n^R|u_m^L\rangle\langle u_m^R|u_n^R\rangle\langle u_n^R|\partial_{k_j} u_n^R\rangle\\
    =&\sum_{m\neq n} \langle-i\partial_{k_i} u_n^R|u_m^L\rangle\langle u_m^R|i\partial_{k_j} u_n^R\rangle-\langle -i\partial_{k_i} u_n^R|u_m^L\rangle\langle u_m^R|u_n^R\rangle\langle u_n^R|i\partial_{k_j} u_n^R\rangle\\
    =&\sum_{m\neq n} (A_{nm}^{RL})_i(A_{mn}^{RR})_j-I_{mn}(A_{nm}^{RL})_i(A_{nn}^{RR})_j
\end{split}
\end{equation*}
where we denote $I_{mn}=\langle u_m^R|u_n^R\rangle$, which arise from the non-orthogonality of the right eigenstates. For a two-band system, the RR QGT simplifies to
$$
Q_{n,ij}^{RR} = A_{nm,i}^{RL} A_{mn,j}^{RR} - I_{mn}A_{nm,i}^{RL} A_{nn,j}^{RR}.
$$

\section{Derivation of Equation of Motion}
In this section, we will present our derivation of the equation of motion presented in the main text. First, we consider a non-Hermitian Hamiltonian under a constant force $\mathbf{F}$ in the real space
\begin{equation*}\tag{S7}
    \tilde{H}(\mathbf{k})=H(\mathbf{k})-\mathbf{F}\cdot\mathbf{r}
\end{equation*}
where $\tilde{H}$ has left and right eigenstates $|\tilde{u}_n^{L/R}\rangle$ and corresponding eigenenergies $\tilde{\epsilon}_n$, and $H(\mathbf{k})$ has left and right eigenstates denoted as $|u_n^{L/R}\rangle$ and corresponding eigenenergies $\epsilon_n$.

We begin by considering a wavepacket in the eigenstate $|\tilde{u}_0^R\rangle$, which has the largest imaginary parts of the eigenenergies (the smallest decaying rate)
\begin{equation*}\tag{S8}
    |W\rangle=\int w(\mathbf{k},t) e^{i\mathbf{k}\cdot\mathbf{r}}|\tilde{u}_0^R\rangle d\mathbf{k}.
\end{equation*}
We then expand $|W\rangle$ to first order using the perturbation theory in Ref. \cite{sternheim1972}
\begin{equation*}\tag{S9}
    |W\rangle=\int w(\mathbf{k},t) e^{i\mathbf{k}\cdot\mathbf{r}}\bigg(|u_0^R\rangle+\frac{\mathbf{F}\cdot\mathbf{A}_{10}^{LR}}{\epsilon_1-\epsilon_0}|u_1^R\rangle\bigg)d\mathbf{k}
\end{equation*}
where the index $1$ denotes the other band in the non-Hermitian two-band system.
From this point forward, we denote $\int_\mathbf{k}=\int d\mathbf{k}$ and $w(\mathbf{k},t)=w$ for brevity.

\subsection{Right-Right Quantum Geometric Tensor as Field-Induced Positional Shift}
To show that the RR QGT describes the field-induced positional shift \cite{gao2014,gao2015WP}, we first calculate the centre-of-mass (COM) position of the wave packet $|W\rangle$, $\mathbf{r}_c=\frac{\langle W|\hat{\mathbf{r}}|W\rangle}{\langle W|W\rangle}$. When we expand the numerator in $\mathbf{r}_c$ to first order perturbation, it can be rewritten as
\begin{equation*}\tag{S10}\label{eq: s10}
    \begin{split}
        \langle W|\hat{\mathbf{r}}|W\rangle=&\int_\mathbf{k} \Bigg[i w^*\partial_\mathbf{k}w\bigg(1+\frac{\mathbf{F}\cdot\mathbf{A}_{10}^{LR}}{\epsilon_1-\epsilon_0}I_{01}+\frac{\mathbf{F}\cdot\mathbf{A}_{01}^{RL}}{\epsilon_1^*-\epsilon_0^*}I_{10}\bigg)+|w|^2\bigg(\langle u_0^R|i\partial_\mathbf{k}u_0^R\rangle+\langle u_1^R|i\partial_\mathbf{k}u_0^R\rangle\frac{\mathbf{F}\cdot\mathbf{A}_{01}^{RL}}{\epsilon_1^*-\epsilon_0^*}\\
        &+\langle u_0^R|i\partial_\mathbf{k} u_1^R\rangle\frac{\mathbf{F}\cdot\mathbf{A}_{10}^{LR}}{\epsilon_1-\epsilon_0}+i\partial_\mathbf{k}\Big(\frac{\mathbf{F}\cdot\mathbf{A}_{10}^{LR}}{\epsilon_1-\epsilon_0}\Big)I_{01}\bigg) \Bigg].
    \end{split}
\end{equation*}
Suppose that the wave packet takes the form $w=|w|e^{i\varphi}$, the first 3 terms in Eq.~(\ref{eq: s10}) can be rewritten as
\begin{equation*}\tag{S11}\label{eq: s11}
    \begin{split}
        \int_\mathbf{k} i w^*\partial_\mathbf{k}w\bigg(1+\frac{\mathbf{F}\cdot\mathbf{A}_{10}^{LR}}{\epsilon_1-\epsilon_0}I_{01}+\frac{\mathbf{F}\cdot\mathbf{A}_{01}^{RL}}{\epsilon_1^*-\epsilon_0^*}I_{10}\bigg)=&\int_\mathbf{k} (-|w|^2\partial_\mathbf{k}\varphi+i|w|\partial_\mathbf{k}|w|)\bigg(1+\frac{\mathbf{F}\cdot\mathbf{A}_{10}^{LR}}{\epsilon_1-\epsilon_0}I_{01}+\frac{\mathbf{F}\cdot\mathbf{A}_{01}^{RL}}{\epsilon_1^*-\epsilon_0^*}I_{10}\bigg)\\
        =&\int_\mathbf{k} |w|^2\Bigg(-\partial_\mathbf{k}\varphi\bigg(1+\frac{\mathbf{F}\cdot\mathbf{A}_{10}^{LR}}{\epsilon_1-\epsilon_0}I_{01}+\frac{\mathbf{F}\cdot\mathbf{A}_{01}^{RL}}{\epsilon_1^*-\epsilon_0^*}I_{10}\bigg)\\
        &-\frac{i}{2}\partial_\mathbf{k}\bigg(\frac{\mathbf{F}\cdot\mathbf{A}_{10}^{LR}}{\epsilon_1-\epsilon_0}I_{01}+\frac{\mathbf{F}\cdot\mathbf{A}_{01}^{RL}}{\epsilon_1^*-\epsilon_0^*}I_{10}\bigg)\Bigg).
    \end{split}
\end{equation*}

We can then plug Eq,~(\ref{eq: s11}) back to Eq.~(\ref{eq: s10}) which yields
    \begin{equation*}\tag{S12}
    \begin{split}
        \langle W|\hat{\mathbf{r}}|W\rangle=&\int_\mathbf{k} |w|^2\bigg(-\partial_\mathbf{k}\varphi\bigg(1+\frac{\mathbf{F}\cdot\mathbf{A}_{10}^{LR}}{\epsilon_1-\epsilon_0}I_{01}+\frac{\mathbf{F}\cdot\mathbf{A}_{01}^{RL}}{\epsilon_1^*-\epsilon_0^*}I_{10}\bigg)-\frac{i}{2}\partial_\mathbf{k}\bigg(\frac{\mathbf{F}\cdot\mathbf{A}_{10}^{LR}}{\epsilon_1-\epsilon_0}I_{01}+\frac{\mathbf{F}\cdot\mathbf{A}_{01}^{RL}}{\epsilon_1^*-\epsilon_0^*}I_{10}\bigg)\\
        &+\langle u_0^R|i\partial_\mathbf{k}u_0^R\rangle+\langle u_1^R|i\partial_\mathbf{k}u_0^R\rangle\frac{\mathbf{F}\cdot\mathbf{A}_{01}^{RL}}{\epsilon_1^*-\epsilon_0^*}+\langle u_0^R|i\partial_\mathbf{k} u_1^R\rangle\frac{\mathbf{F}\cdot\mathbf{A}_{10}^{LR}}{\epsilon_1-\epsilon_0}+i\partial_\mathbf{k}\Big(\frac{\mathbf{F}\cdot\mathbf{A}_{10}^{LR}}{\epsilon_1-\epsilon_0}\Big)I_{01}\bigg)\\
        =&\int_\mathbf{k} |w|^2\bigg(-\partial_\mathbf{k}\varphi\bigg(1+\frac{\mathbf{F}\cdot\mathbf{A}_{10}^{LR}}{\epsilon_1-\epsilon_0}I_{01}+\frac{\mathbf{F}\cdot\mathbf{A}_{01}^{RL}}{\epsilon_1^*-\epsilon_0^*}I_{10}\bigg)+\frac{i}{2}\Big(\partial_\mathbf{k}\frac{\mathbf{F}\cdot\mathbf{A}_{10}^{LR}}{\epsilon_1-\epsilon_0}\Big)I_{01}-\frac{i}{2}\frac{\mathbf{F}\cdot\mathbf{A}_{10}^{LR}}{\epsilon_1-\epsilon_0}\partial_\mathbf{k}I_{01}\\
        &-\frac{i}{2}\partial_\mathbf{k}\Big(\frac{\mathbf{F}\cdot\mathbf{A}_{01}^{RL}}{\epsilon_1^*-\epsilon_0^*}\Big)I_{10}-\frac{i}{2}\frac{\mathbf{F}\cdot\mathbf{A}_{01}^{RL}}{\epsilon_1^*-\epsilon_0^*}\partial_\mathbf{k}I_{10}+\langle u_0^R|i\partial_\mathbf{k}u_0^R\rangle+\langle u_1^R|i\partial_\mathbf{k}u_0^R\rangle\frac{\mathbf{F}\cdot\mathbf{A}_{01}^{RL}}{\epsilon_1^*-\epsilon_0^*}+\langle u_0^R|i\partial_\mathbf{k} u_1^R\rangle\frac{\mathbf{F}\cdot\mathbf{A}_{10}^{LR}}{\epsilon_1-\epsilon_0}\bigg)
    \end{split}
    \end{equation*}

Suppose that the wave packet is narrow, we can approximate $|w|^2\approx \delta(\mathbf{k}-\mathbf{k}_c)$, where $\mathbf{k}_c$ is the COM momentum. The equation can be rewritten as
    \begin{equation*}\tag{S13}\label{eq: s13}
        \begin{split}
            \langle W|\hat{\mathbf{r}}|W\rangle=&-\partial_\mathbf{k}\varphi+\mathbf{A}_{00}^{RR}+\frac{\mathbf{F}\cdot\mathbf{A}_{10}^{LR}}{\epsilon_1-\epsilon_0}\bigg(-\partial_\mathbf{k}\varphi I_{01}+\mathbf{A}_{01}^{RR}-\frac{i}{2}\partial_\mathbf{k}I_{01}\bigg)+\frac{\mathbf{F}\cdot\mathbf{A}_{01}^{RL}}{\epsilon_1^*-\epsilon_0^*}\bigg(-\partial_\mathbf{k}\varphi I_{10}+\mathbf{A}_{10}^{RR}-\frac{i}{2}\partial_\mathbf{k}I_{10}\bigg)\\
            &+\frac{i}{2}I_{01}\Big(\partial_\mathbf{k}\frac{\mathbf{F}\cdot\mathbf{A}_{10}^{LR}}{\epsilon_1-\epsilon_0}\Big)-\frac{i}{2}I_{10}\Big(\partial_\mathbf{k}\frac{\mathbf{F}\cdot\mathbf{A}_{01}^{RL}}{\epsilon_1^*-\epsilon_0^*}\Big)|_{\mathbf{k}=\mathbf{k}_c}\\
            =&-\partial_\mathbf{k}\varphi+\mathbf{A}_{00}^{RR}+\frac{\mathbf{F}\cdot\mathbf{A}_{10}^{LR}}{\epsilon_1-\epsilon_0}\bigg(-\partial_\mathbf{k}\varphi I_{01}+\mathbf{A}_{01}^{RR}-i\partial_\mathbf{k}I_{01}\bigg)+\frac{\mathbf{F}\cdot\mathbf{A}_{01}^{RL}}{\epsilon_1^*-\epsilon_0^*}\bigg(-\partial_\mathbf{k}\varphi I_{10}+\mathbf{A}_{10}^{RR}\bigg)\\
            &+\frac{i}{2}\partial_\mathbf{k}\Big(I_{01}\frac{\mathbf{F}\cdot\mathbf{A}_{10}^{LR}}{\epsilon_1-\epsilon_0}\Big)-\frac{i}{2}\partial_\mathbf{k}\Big(I_{10}\frac{\mathbf{F}\cdot\mathbf{A}_{01}^{RL}}{\epsilon_1^*-\epsilon_0^*}\Big)|_{\mathbf{k}=\mathbf{k}_c}.
        \end{split}
    \end{equation*}

We then also note that the term $\mathbf{A}_{01}^{RR}-i\partial_\mathbf{k}I_{01}$ can be identified as $(\mathbf{A}_{10}^{RR})^*$ since
\begin{equation*}\tag{S14}
\begin{split}
    -i\partial_\mathbf{k}I_{01}&=\langle-i\partial_\mathbf{k}u_0^R|u_1^R\rangle-\langle u_0^R|i\partial_\mathbf{k}u_1^R\rangle\\
    &=(\mathbf{A}_{10}^{RR})^*-\mathbf{A}_{01}^{RR}\\
    (\mathbf{A}_{10}^{RR})^*&=\mathbf{A}_{01}^{RR}-i\partial_\mathbf{k}I_{01}.
\end{split}
\end{equation*}

 Additionally, the term $I_{01}\mathbf{A}_{10}^{LR}$ can also be identified as $\mathbf{A}_{00}^{RR}-\mathbf{A}_{00}^{LR}$ since
\begin{equation*}\tag{S15}
    \begin{split}
        \mathbf{A}_{00}^{RR}&=\langle u_0^R|i\partial_\mathbf{k}u_0^R\rangle\\
        &=\langle u_0^R|u_0^R \rangle\langle u_0^L|i\partial_\mathbf{k}u_0^R\rangle+\langle u_0^R|u_1^R \rangle\langle u_1^L|i\partial_\mathbf{k}u_0^R\rangle\\
        &=\mathbf{A}_{00}^{LR}+I_{01}\mathbf{A}_{10}^{LR}\\
        I_{01}\mathbf{A}_{10}^{LR}&=\mathbf{A}_{00}^{RR}-\mathbf{A}_{00}^{LR}.
    \end{split}
\end{equation*}

We then expand the denominator in $\mathbf{r}_c$ to first order in $\mathbf{F}$ using geometric series
\begin{equation*}\tag{S16}\label{eq: s16}
    \begin{split}
        \frac{1}{\langle W|W \rangle}
        &=\bigg(1+\frac{\mathbf{F}\cdot\mathbf{A}_{10}^{LR}}{\epsilon_1-\epsilon_0}I_{01}+\frac{\mathbf{F}\cdot\mathbf{A}_{01}^{RL}}{\epsilon_1^*-\epsilon_0^*}I_{10}|_{\mathbf{k}=\mathbf{k}_c}\bigg)^{-1}\\
        &\approx \bigg(1-\frac{\mathbf{F}\cdot\mathbf{A}_{10}^{LR}}{\epsilon_1-\epsilon_0}I_{01}-\frac{\mathbf{F}\cdot\mathbf{A}_{01}^{RL}}{\epsilon_1^*-\epsilon_0^*}I_{10}\bigg)|_{\mathbf{k}=\mathbf{k}_c}.
    \end{split}
\end{equation*}

We can finally combine $\langle W|\hat{\mathbf{r}}|W\rangle$ from Eq.~(\ref{eq: s13}) and $\langle W|W\rangle$ from Eq.~(\ref{eq: s16}) and obtain $\mathbf{r}_c$ up to the first order perturbation
    \begin{equation*}\tag{S17}
        \begin{split}
            \mathbf{r}_c=&\frac{\langle W|\hat{\mathbf{r}}|W\rangle}{\langle W|W\rangle}\\
            =&\Big(-\partial_\mathbf{k}\varphi+\mathbf{A}_{00}^{RR}\Big)\bigg(1-\frac{\mathbf{F}\cdot\mathbf{A}_{10}^{LR}}{\epsilon_1-\epsilon_0}I_{01}-\frac{\mathbf{F}\cdot\mathbf{A}_{01}^{RL}}{\epsilon_1^*-\epsilon_0^*}I_{10}\bigg)+\frac{\mathbf{F}\cdot\mathbf{A}_{10}^{LR}}{\epsilon_1-\epsilon_0}\bigg(-\partial_\mathbf{k}\varphi I_{01}+(\mathbf{A}_{10}^{RR})^*\bigg)\\
            &+\frac{\mathbf{F}\cdot\mathbf{A}_{01}^{RL}}{\epsilon_1^*-\epsilon_0^*}\bigg(-\partial_\mathbf{k}\varphi I_{10}+\mathbf{A}_{10}^{RR}\bigg)+\frac{i}{2}\partial_\mathbf{k}\Big(\frac{\mathbf{F}\cdot(\mathbf{A}_{00}^{RR}-\mathbf{A}_{00}^{LR})}{\epsilon_1-\epsilon_0}\Big)-\frac{i}{2}\partial_\mathbf{k}\Big(\frac{\mathbf{F}\cdot(\mathbf{A}_{00}^{RR}-\mathbf{A}_{00}^{LR})^*}{\epsilon_1^*-\epsilon_0^*}\Big)|_{\mathbf{k}=\mathbf{k}_c}\\
            =&-\partial_\mathbf{k}\varphi+\mathbf{A}_{00}^{RR}+\frac{\mathbf{F}\cdot\mathbf{A}_{10}^{LR}}{\epsilon_1-\epsilon_0}\bigg((\mathbf{A}_{10}^{RR})^*-I_{01}\mathbf{A}_{00}^{RR}\bigg)+\frac{\mathbf{F}\cdot\mathbf{A}_{01}^{RL}}{\epsilon_1^*-\epsilon_0^*}\bigg(\mathbf{A}_{10}^{RR}-I_{10}\mathbf{A}_{00}^{RR}\bigg)\\
            &+\frac{i}{2}\partial_\mathbf{k}\Big(\frac{\mathbf{F}\cdot(\mathbf{A}_{00}^{RR}-\mathbf{A}_{00}^{LR})}{\epsilon_1-\epsilon_0}\Big)-\frac{i}{2}\partial_\mathbf{k}\Big(\frac{\mathbf{F}\cdot(\mathbf{A}_{00}^{RR}-\mathbf{A}_{00}^{LR})^*}{\epsilon_1^*-\epsilon_0^*}\Big)|_{\mathbf{k}=\mathbf{k}_c}\\
            =&-\partial_\mathbf{k}\varphi+\mathbf{A}_{00}^{RR}+2\operatorname{Re}\bigg[\frac{\mathbf{F}\cdot (Q_{0}^{RR})^*}{\epsilon_1-\epsilon_0}\bigg]-\operatorname{Im}\bigg[\partial_\mathbf{k}\Big(\frac{\mathbf{F}\cdot(\mathbf{A}_{00}^{RR}-\mathbf{A}_{00}^{LR})}{\epsilon_1-\epsilon_0}\Big)\bigg]|_{\mathbf{k}=\mathbf{k}_c}.
        \end{split}
    \end{equation*}

Since the RR QGT is a Hermitian tensor $(Q_{0,ij}^{RR})^*=Q_{0,ji}^{RR}$, we can rewrite the third term using the identity $\mathbf{F}\cdot (Q_0^{RR})^*=Q_0^{RR}\cdot\mathbf{F}$, which yields the result

\begin{equation*}\tag{S18}\label{eq: r}
    \mathbf{r}_c=-\partial_\mathbf{k}\varphi+\mathbf{A}_{00}^{RR}+2\operatorname{Re}\bigg[\frac{Q_{0}^{RR}\cdot\mathbf{F}}{\epsilon_1-\epsilon_0}\bigg]-\operatorname{Im}\bigg[\partial_\mathbf{k}\Big(\frac{\mathbf{F}\cdot(\mathbf{A}_{00}^{RR}-\mathbf{A}_{00}^{LR})}{\epsilon_1-\epsilon_0}\Big)\bigg]|_{\mathbf{k}=\mathbf{k}_c}
\end{equation*}
which shows that the RR QGT together with the anomalous Berry connection $\mathbf{A}_{00}^{RR}-\mathbf{A}_{00}^{LR}$ describe the field-induced positional shift.

\subsection{Left-Right Quantum Geometric Tensor as Correction to Phase}
To derive how the LR QGT appears in the equation of motion, we follow the same formalism presented in Refs. \cite{xu2017,wang2022}. We start from the time-dependent Schrodinger equation of the wave packet
\begin{equation*}\tag{S19}
    i\hbar\partial_t\Big(w e^{i\mathbf{k}\cdot\mathbf{r}}|\tilde{u}_0^R\rangle\Big)=(H-\mathbf{F}\cdot\mathbf{r})\Big(w e^{i\mathbf{k}\cdot\mathbf{r}}|\tilde{u}_0^R\rangle\Big).
\end{equation*}
We then multiply both sides with $\langle\tilde{\psi}|=\sum_n \tilde{c}_n^L\langle\tilde{u}_n^L|$, where $|\tilde{u}_n^L\rangle$ denote the $n$-th left eigenstate of the Hamiltonian $\tilde{H}=H-\mathbf{F}\cdot\mathbf{r}$,
\begin{equation*}\tag{S20}\label{eq: s20}
\begin{split}
    \sum_n \tilde{c}_n^L\langle\tilde{u}_n^L|i\hbar\partial_tw e^{i\mathbf{k}\cdot\mathbf{r}}|\tilde{u}_0^R\rangle&=\sum_n \tilde{c}_n^L\langle\tilde{u}_n^L|(H-\mathbf{F}\cdot\mathbf{r})w e^{i\mathbf{k}\cdot\mathbf{r}}|\tilde{u}_0^R\rangle.
\end{split}
\end{equation*}
Note that for any $\tilde{c}_n^L$, all terms proportional to $\sum_{n\neq0}\tilde{c}_n^L\langle\tilde{u}_n^L|$ will vanish since
\begin{equation*}\tag{S21}
    \begin{split}
        \sum_{n\neq0} \tilde{c}_n^L\langle\tilde{u}_n^L|i\hbar\partial_tw e^{i\mathbf{k}\cdot\mathbf{r}}|\tilde{u}_0^R\rangle&=\sum_{n\neq0} \tilde{c}_n^Lw e^{i\mathbf{k}\cdot\mathbf{r}}\langle\tilde{u}_n^L|(H-\mathbf{F}\cdot\mathbf{r})|\tilde{u}_0^R\rangle\\
        &=\sum_{n\neq0}\tilde{c}_n^Lw e^{i\mathbf{k}\cdot\mathbf{r}}\tilde{\epsilon}_n\langle\tilde{u}_n^L|\tilde{u}_0^R\rangle\\
        &=0,
    \end{split}
\end{equation*}
and only the terms proportional to $\langle\tilde{u}_0^L|$ survive.

We can then expand both sides of Eq.~(\ref{eq: s20}) and remove the overall factor of $\tilde{c}_0^L$ as
    \begin{equation*}\tag{S22}\label{eq: s22}
        \begin{split}
            \langle\tilde{u}_0^L|i\hbar\partial_t\Big(w e^{i\mathbf{k}\cdot\mathbf{r}}|\tilde{u}_0^R\rangle\Big)=&\bigg(\langle u_0^L|+\frac{\mathbf{F}\cdot\mathbf{A}_{01}^{LR}}{\epsilon_1-\epsilon_0}\langle u_1^L|\bigg)i\hbar\partial_t\bigg(w e^{i\mathbf{k}\cdot\mathbf{r}}\Big(|u_0^R\rangle+\frac{\mathbf{F}\cdot\mathbf{A}_{10}^{LR}}{\epsilon_1-\epsilon_0}|u_1^R\rangle\Big)\bigg)\\
            =&\bigg(\langle u_0^L|+\frac{\mathbf{F}\cdot\mathbf{A}_{01}^{LR}}{\epsilon_1-\epsilon_0}\langle u_1^L|\bigg)\bigg(i\hbar(\partial_tw)e^{i\mathbf{k}\cdot\mathbf{r}}|u_0^R\rangle-\hbar(\partial_t\mathbf{k}\cdot\mathbf{r})e^{i\mathbf{k}\cdot\mathbf{r}}|u_0^R\rangle+i\hbar we^{i\mathbf{k}\cdot\mathbf{r}}\partial_t|u_0^R\rangle\\
            &+i\hbar(\partial_tw)e^{i\mathbf{k}\cdot\mathbf{r}}\frac{\mathbf{F}\cdot\mathbf{A}_{10}^{LR}}{\epsilon_1-\epsilon_0}|u_1^R\rangle-\hbar(\partial_t\mathbf{k}\cdot\mathbf{r})e^{i\mathbf{k}\cdot\mathbf{r}}\frac{\mathbf{F}\cdot\mathbf{A}_{10}^{LR}}{\epsilon_1-\epsilon_0}|u_1^R\rangle+i\hbar we^{i\mathbf{k}\cdot\mathbf{r}}\Big(\partial_t\frac{\mathbf{F}\cdot\mathbf{A}_{10}^{LR}}{\epsilon_1-\epsilon_0}\Big)|u_1^R\rangle\\
            &+i\hbar we^{i\mathbf{k}\cdot\mathbf{r}}\frac{\mathbf{F}\cdot\mathbf{A}_{10}^{LR}}{\epsilon_1-\epsilon_0}\partial_t|u_1^R\rangle\bigg)\\
           \approx&i\hbar(\partial_tw)e^{i\mathbf{k}\cdot\mathbf{r}}-\hbar w(\partial_t\mathbf{k}\cdot\mathbf{r})e^{i\mathbf{k}\cdot\mathbf{r}}+i\hbar we^{i\mathbf{k}\cdot\mathbf{r}}\langle u_0^L|\partial_t|u_0^R\rangle+i\hbar we^{i\mathbf{k}\cdot\mathbf{r}}\frac{\mathbf{F}\cdot\mathbf{A}_{10}^{LR}}{\epsilon_1-\epsilon_0}\langle u_0^L|\partial_t|u_1^R\rangle\\
           &+i\hbar we^{i\mathbf{k}\cdot\mathbf{r}}\frac{\mathbf{F}\cdot\mathbf{A}_{01}^{LR}}{\epsilon_1-\epsilon_0}\langle u_1^L|\partial_t|u_0^R\rangle
        \end{split}
    \end{equation*}
    \begin{equation*}\tag{S23}\label{eq: s23}
    \begin{split}
        \langle\tilde{u}_0^L|(H-\mathbf{F}\cdot\mathbf{r})\Big(w e^{i\mathbf{k}\cdot\mathbf{r}}|\tilde{u}_0^R\rangle\Big)=&\bigg(\langle u_0^L|+\frac{\mathbf{F}\cdot\mathbf{A}_{01}^{LR}}{\epsilon_1-\epsilon_0}\langle u_1^L|\bigg)(H-\mathbf{F}\cdot\mathbf{r})\bigg(w e^{i\mathbf{k}\cdot\mathbf{r}}\Big(|u_0^R\rangle+\frac{\mathbf{F}\cdot\mathbf{A}_{10}^{LR}}{\epsilon_1-\epsilon_0}|u_1^R\rangle\Big)\bigg)\\
        =&\bigg(\langle u_0^L|+\frac{\mathbf{F}\cdot\mathbf{A}_{01}^{LR}}{\epsilon_1-\epsilon_0}\langle u_1^L|\bigg)\bigg(w e^{i\mathbf{k}\cdot\mathbf{r}}\epsilon_0|u_0^R\rangle-w e^{i\mathbf{k}\cdot\mathbf{r}}(\mathbf{F}\cdot\mathbf{r})|u_0^R\rangle\\
        &+w e^{i\mathbf{k}\cdot\mathbf{r}}\frac{\mathbf{F}\cdot\mathbf{A}_{10}^{LR}}{\epsilon_1-\epsilon_0}\epsilon_1|u_1^R\rangle-w e^{i\mathbf{k}\cdot\mathbf{r}}\frac{\mathbf{F}\cdot\mathbf{A}_{10}^{LR}}{\epsilon_1-\epsilon_0}(\mathbf{F}\cdot\mathbf{r})|u_1^R\rangle\bigg)\\
        \approx&w e^{i\mathbf{k}\cdot\mathbf{r}}\epsilon_0-w e^{i\mathbf{k}\cdot\mathbf{r}}(\mathbf{F}\cdot\mathbf{r})
    \end{split}
    \end{equation*}
where we discard all the higher-order terms and only keep up to first order perturbation

Assuming the wave packet is infinitely narrow in momentum space, the dynamics of its centre-of-mass momentum is governed by the external force 
$\dot{\mathbf{k}}_c=\mathbf{F}/\hbar$, which allow us to rewrite the right hand side of the Eq.~(\ref{eq: s22}) as

    \begin{equation*}\tag{S24}\label{eq: s24}
        \begin{split}
            \langle\tilde{u}_0^L|i\hbar\partial_t\Big(w e^{i\mathbf{k}\cdot\mathbf{r}}|\tilde{u}_0^R\rangle\Big)=&i\hbar(\partial_tw)e^{i\mathbf{k}\cdot\mathbf{r}}-w(\mathbf{F}\cdot\mathbf{r})e^{i\mathbf{k}\cdot\mathbf{r}}+we^{i\mathbf{k}\cdot\mathbf{r}}\mathbf{F}\cdot\langle u_0^L|i\partial_\mathbf{k}|u_0^R\rangle\\
            &+ we^{i\mathbf{k}\cdot\mathbf{r}}\frac{\mathbf{F}\cdot\mathbf{A}_{10}^{LR}}{\epsilon_1-\epsilon_0}\mathbf{F}\cdot\langle u_0^L|i\partial_\mathbf{k}|u_1^R\rangle+we^{i\mathbf{k}\cdot\mathbf{r}}\frac{\mathbf{F}\cdot\mathbf{A}_{01}^{LR}}{\epsilon_1-\epsilon_0}\mathbf{F}\cdot\langle u_1^L|i\partial_\mathbf{k}|u_0^R\rangle\\
            =&i\hbar(\partial_tw)e^{i\mathbf{k}\cdot\mathbf{r}}-w(\mathbf{F}\cdot\mathbf{r})e^{i\mathbf{k}\cdot\mathbf{r}}+we^{i\mathbf{k}\cdot\mathbf{r}}(\mathbf{F}\cdot\mathbf{A}_{00}^{LR})\\
            &+ we^{i\mathbf{k}\cdot\mathbf{r}}\frac{\mathbf{F}\cdot\mathbf{A}_{10}^{LR}}{\epsilon_1-\epsilon_0}(\mathbf{F}\cdot\mathbf{A}_{01}^{LR})+we^{i\mathbf{k}\cdot\mathbf{r}}\frac{\mathbf{F}\cdot\mathbf{A}_{01}^{LR}}{\epsilon_1-\epsilon_0}(\mathbf{F}\cdot\mathbf{A}_{10}^{LR}).
        \end{split}
    \end{equation*}

After equating Eq.~(\ref{eq: s23}) with Eq.~(\ref{eq: s24}), we can remove the overall factor of $e^{i\mathbf{k}\cdot\mathbf{r}}$ and remove the terms that appear on both sides, yielding
    \begin{equation*}\tag{S25}
        \begin{split}
       w \epsilon_0=&i\hbar(\partial_tw)+w(\mathbf{F}\cdot\mathbf{A}_{00}^{LR})+ w\frac{\mathbf{F}\cdot\mathbf{A}_{10}^{LR}}{\epsilon_1-\epsilon_0}(\mathbf{F}\cdot\mathbf{A}_{01}^{LR})+w\frac{\mathbf{F}\cdot\mathbf{A}_{01}^{LR}}{\epsilon_1-\epsilon_0}(\mathbf{F}\cdot\mathbf{A}_{10}^{LR}),
        \end{split}
    \end{equation*}
after rearrangement, this gives a differential equation of $w$
    \begin{equation*}\tag{S26}\label{eq: s26}
        \begin{split}
            \partial_tw&=-\frac{i}{\hbar}w\bigg( \epsilon_0-(\mathbf{F}\cdot\mathbf{A}_{00}^{LR})- \frac{\mathbf{F}\cdot\mathbf{A}_{10}^{LR}}{\epsilon_1-\epsilon_0}(\mathbf{F}\cdot\mathbf{A}_{01}^{LR})-\frac{\mathbf{F}\cdot\mathbf{A}_{01}^{LR}}{\epsilon_1-\epsilon_0}(\mathbf{F}\cdot\mathbf{A}_{10}^{LR})\bigg)\\
            &=-\frac{i}{\hbar}w\bigg( \epsilon_0-(\mathbf{F}\cdot\mathbf{A}_{00}^{LR})- \frac{2(\mathbf{F}\cdot g^{LR}\cdot\mathbf{F})}{\epsilon_1-\epsilon_0}\bigg)
        \end{split}
    \end{equation*}
where we use the identity of the LR quantum metric from the previous section, we also abbreviated the index in $g^{LR}$ since in a two-band system, the LR quantum metric tensors of the two bands are the same. Eq.~(\ref{eq: s26}) has a general solution of $w$ in the form of
\begin{equation*}\tag{S27}
    w\propto\operatorname{exp}\bigg[-\frac{i}{\hbar}\int_{t_0}^{t}\bigg( \epsilon_0-\mathbf{F}\cdot\bigg(\mathbf{A}_{00}^{LR}+ \frac{2g^{LR}\cdot\mathbf{F}}{\epsilon_1-\epsilon_0}\bigg) \bigg)|_{\mathbf{k}=\mathbf{k}_c}dt'\bigg],
\end{equation*}
which yields the correction to the phase of the wave packet as
\begin{equation*}\tag{S28}\label{eq: phi}
    \varphi=-\frac{1}{\hbar}\int_{t_0}^{t}\operatorname{Re}\bigg[ \epsilon_0-\mathbf{F}\cdot\bigg(\mathbf{A}_{00}^{LR}+\frac{2g^{LR}\cdot\mathbf{F}}{\epsilon_1-\epsilon_0}\bigg)|_{\mathbf{k}=\mathbf{k}_c}\bigg]dt'.
\end{equation*}
Finally, plugging Eq.~(\ref{eq: phi}) into Eq.~(\ref{eq: r}) and taking the first derivative, we  derive the group velocity under first order perturbation as
\begin{equation*}\tag{S29}\label{eq: main result}
    \begin{split}
        (\dot{\mathbf{r}}_c)_i=&\partial_t\bigg(-\partial_{k_i}\varphi+(A_{00}^{RR})_i+2\operatorname{Re}\bigg[\frac{(Q_{0}^{RR}\cdot\mathbf{F})_i}{\epsilon_1-\epsilon_0}\bigg]-\operatorname{Im}\bigg[\partial_{k_i}\Big(\frac{\mathbf{F}\cdot(\mathbf{A}_{00}^{RR}-\mathbf{A}_{00}^{LR})}{\epsilon_1-\epsilon_0}\Big)\bigg]|_{\mathbf{k}=\mathbf{k}_c}\bigg)\\
        =&\frac{1}{\hbar}\partial_{k_i}\operatorname{Re}\bigg[ \epsilon_0-\sum_j F_j\bigg((A_{00}^{LR})_j+\frac{2(g^{LR}\cdot\mathbf{F})_j}{\epsilon_1-\epsilon_0}\bigg)\bigg]\\
        &+\partial_t\bigg((A_{00}^{RR})_i+2\operatorname{Re}\bigg[\frac{(Q_{0}^{RR}\cdot\mathbf{F})_i}{\epsilon_1-\epsilon_0}\bigg]-\operatorname{Im}\bigg[\partial_{k_i}\Big(\frac{\mathbf{F}\cdot(\mathbf{A}_{00}^{RR}-\mathbf{A}_{00}^{LR})}{\epsilon_1-\epsilon_0}\Big)\bigg]\bigg)|_{\mathbf{k}=\mathbf{k}_c}\\
        =&\frac{1}{\hbar}\partial_{k_i}\operatorname{Re}\bigg[ \epsilon_0-\sum_j F_j\bigg((A_{00}^{LR})_j+\frac{2(g^{LR}\cdot\mathbf{F})_j}{\epsilon_1-\epsilon_0}\bigg)\bigg]\\
        &+\frac{1}{\hbar}\sum_j F_j\partial_{k_j}\bigg((A_{00}^{RR})_i+2\operatorname{Re}\bigg[\frac{(Q_{0}^{RR}\cdot\mathbf{F})_i}{\epsilon_1-\epsilon_0}\bigg]-\operatorname{Im}\bigg[\partial_{k_i}\Big(\frac{\mathbf{F}\cdot(\mathbf{A}_{00}^{RR}-\mathbf{A}_{00}^{LR})}{\epsilon_1-\epsilon_0}\Big)\bigg]\bigg)|_{\mathbf{k}=\mathbf{k}_c}
    \end{split}
\end{equation*}

which is the main result presented in the main text.

\section{Real and Imaginary Parts of Quantum Geometric Tensors}
We also note that the real and imaginary parts in the terms in Eq.~(\ref{eq: main result}) can be expanded as
\begin{equation*}\label{eq: re and im part}\tag{S30}
    \begin{split}
        \operatorname{Re}\Big[\frac{2(Q_0^{RR}\cdot \mathbf{F})}{\Delta\epsilon}\Big]=&\frac{\operatorname{Re}[\Delta\epsilon](2g_{0}^{RR}\cdot \mathbf{F})}{|\Delta\epsilon|^2}-\frac{\operatorname{Im}[\Delta\epsilon](\mathbf{F}\times\mathbf{\Omega}_0^{RR})}{|\Delta\epsilon|^2}\\
        \operatorname{Im}\Big[\frac{\mathbf{F}\cdot(\mathbf{A}_{00}^{RR}-\mathbf{A}_{00}^{LR})}{\Delta\epsilon}\Big]=&-\frac{\operatorname{Re}[\Delta\epsilon]\operatorname{Im}[\mathbf{F}\cdot\mathbf{A}_{00}^{LR}]}{|\Delta\epsilon|^2}-\frac{\operatorname{Im}[\Delta\epsilon]\operatorname{Re}[\mathbf{F}\cdot(\mathbf{A}_{00}^{RR}-\mathbf{A}_{00}^{LR})]}{|\Delta\epsilon|^2}\\
        \operatorname{Re}\Big[\frac{2(g_0^{LR}\cdot \mathbf{F})}{\Delta\epsilon}\Big]=&\frac{\operatorname{Re}[\Delta\epsilon]\Big(2\operatorname{Re}[g_0^{LR}]\cdot\mathbf{F}\Big)}{|\Delta\epsilon|^2}+\frac{\operatorname{Im}[\Delta\epsilon]\Big(2\operatorname{Im}[g_0^{LR}]\cdot\mathbf{F}\Big)}{|\Delta\epsilon|^2}.
    \end{split}
\end{equation*}
where we denote $\Delta\epsilon=\epsilon_1-\epsilon_0$, and $\operatorname{Im}[\mathbf{A}_{00}^{RR}]$ vanishes since $\mathbf{A}_{00}^{RR}$ is real-valued. This suggest that the RR Berry curvature also plays a role in the correction to the non-Hermitian anomalous Berry connection. Furthermore, it shows that both the real and imaginary parts of the LR QMT are needed to be taken account for in the equation of motion.

To see how each term manifests in the dynamics, we rewrite the equation of motion in the main results using the forms in Eqs.~(\ref{eq: re and im part})
\begin{equation*}\label{eq: re and im eom}\tag{S31}
    \begin{split}
        \hbar\dot{\mathbf{r}}_c=&\nabla_\mathbf{k}\operatorname{Re}[\epsilon_0+\mathbf{F}\cdot(\mathbf{A}_{00}^{RR}-\mathbf{A}_{00}^{LR})]\\
        &+\nabla_\mathbf{k}\Bigg(\mathbf{F}\cdot\Bigg(\frac{\operatorname{Re}[\Delta\epsilon]\Big(2(g_0^{RR}-\operatorname{Re}[g_0^{LR}])\cdot\mathbf{F}\Big)}{|\Delta\epsilon|^2}-\frac{\operatorname{Im}[\Delta\epsilon]\Big(2\operatorname{Im}[g_0^{LR}\cdot\mathbf{F}]\Big)}{|\Delta\epsilon|^2}\Bigg)\\
        &+\mathbf{F}\cdot\nabla_\mathbf{k}\Bigg(\frac{\operatorname{Re}[\Delta\epsilon]\operatorname{Im}[\mathbf{F}\cdot\mathbf{A}_{00}^{LR}]}{|\Delta\epsilon|^2}+\frac{\operatorname{Im}[\Delta\epsilon]\operatorname{Re}[\mathbf{F}\cdot(\mathbf{A}_{00}^{RR}-\mathbf{A}_{00}^{LR})]}{|\Delta\epsilon|^2}\Bigg)\Bigg)\\
        &-\mathbf{F}\times\Bigg(\mathbf{\Omega}_0^{RR}+\nabla_\mathbf{k}\times\frac{\operatorname{Re}[\Delta\epsilon](2g_0^{RR}\cdot\mathbf{F})}{|\Delta\epsilon|^2}+(\mathbf{F}\cdot\nabla_\mathbf{k})\frac{\operatorname{Im}[\Delta\epsilon]\mathbf{\Omega}_0^{RR}}{|\Delta\epsilon|^2}\Bigg).
    \end{split}
\end{equation*}

$$
2\Bigg(\frac{\operatorname{Re}[\Delta\epsilon]\left(g_0^{RR}-\operatorname{Re}[g_0^{LR}]\right)}{|\Delta\epsilon|^2}-\frac{\operatorname{Im}[\Delta\epsilon]\left(\operatorname{Im}[g_0^{LR}]\right)}{|\Delta\epsilon|^2}\Bigg)\cdot\mathbf{F}
$$

From Eq.~(\ref{eq: re and im eom}), we can see that the real and imaginary parts of the Berry connections together with the LR and the RR quantum metric describe the correction to the Berry conections, while the RR quantum metric and the RR Berry curvature describe the correction to the anomalous Hall drift. Here, we use the identity $\nabla_\mathbf{k}\times(-\mathbf{F}\times\mathbf{\Omega}_0^{RR})=\mathbf{F}\cdot\partial_\mathbf{k}\mathbf{\Omega}_0^{RR}$, which holds in 2D systems when $\mathbf{\Omega}_0^{RR}=(0,0,\Omega_0^{z,RR})$.

For the special case presented in the main text, the equation becomes:
\begin{equation*}\tag{S32}
        \begin{split}
            v_x=&\frac{1}{\hbar}\partial_{k_x}\operatorname{Re}\Big[F_y\Big(\tilde{A}_{00,y}^{RR}-\tilde{A}_{00,y}^{LR}\Big)\Big]-\frac{F_y}{\hbar}\tilde{\Omega}_{0,xy}^{RR}\\
            v_y=&\frac{1}{\hbar}\partial_{k_y}\operatorname{Re}\Big[\epsilon_0+F_y\Big(\tilde{A}_{00,y}^{RR}-\tilde{A}_{00,y}^{LR}\Big)\Big].
        \end{split}
    \end{equation*}
where
\begin{widetext}
    \begin{equation}\tag{S33}
        \begin{split}
            \tilde{A}_{00,y}^{RR}-\tilde{A}_{00,y}^{LR}=&\frac{\operatorname{Re}[\Delta\epsilon](2(g_{0,yy}^{RR}-\operatorname{Re}[g_{0,yy}^{LR}])F_y)-\operatorname{Im}[\Delta\epsilon](2\operatorname{Im}[g_{0,yy}^{LR}]F_y)}{|\Delta\epsilon|^2}\\
            &+\partial_{k_y}\bigg(\frac{\operatorname{Re}[\Delta\epsilon]\operatorname{Im}[A_{00,y}^{LR}]F_y+\operatorname{Im}[\Delta\epsilon]\operatorname{Re}[(A_{00,y}^{RR}-A_{00,y}^{LR})]F_y}{|\Delta\epsilon|^2}\bigg)\\
            \tilde{\Omega}_{0,xy}^{RR}=&\Omega_{0,xy}^{RR}+F_y\Bigg(\partial_{k_x}\bigg(\frac{\operatorname{Re}[\Delta\epsilon]g_{0,yy}^{RR}}{|\Delta\epsilon|^2}\bigg)-\partial_{k_y}\bigg(\frac{\operatorname{Re}[\Delta\epsilon]g_{0,xy}^{RR}}{|\Delta\epsilon|^2}\bigg)\Bigg)+F_y\partial_{k_y}\bigg(\frac{\operatorname{Im}[\Delta\epsilon]\Omega_{0,xy}^{RR}}{|\Delta\epsilon|^2}\bigg).
        \end{split}
    \end{equation}
\end{widetext}
Both the RR QGT and the LR QMT appears in the equation of motion, despite that the perturbation theory is defined with respects to the "left-right" formalism and that the centre-of-mass position of the wave packet is calculated using the "right-right" formalism.

\section{Energy eigenvalues and eigenvectors}
The corresponding complex eigenenergies of the polariton Hamiltonian presented in the main text
are
\begin{equation}\tag{S34}
\begin{split}
    \epsilon_\pm&=E_0+\frac{\hbar^2 k^2}{2m}-i\gamma_0-i\gamma_2 k^2-i\gamma_4 k^4\pm\epsilon\\
    \epsilon&=\sqrt{G_x^2+G_y^2+G_z^2}\\
    &=\sqrt{\tilde\alpha^2+2\tilde\alpha\tilde\beta(k_x^2-k_y^2)+\tilde\beta^2k^4+\Delta_z^2}
\end{split}
\end{equation}
and the left and right eigenstates are
\begin{equation}\tag{S35}
    \begin{split}
        |u_\pm^R\rangle&=\frac{1}{\sqrt{|G_z\pm\epsilon|^2+|G_x+iG_y|^2}}\begin{pmatrix} G_z\pm\epsilon \\G_x+iG_y \end{pmatrix}\\
        \langle u_\pm^L|&=\frac{\sqrt{|G_z\pm\epsilon|^2+|G_x+iG_y|^2}}{(G_z\pm\epsilon)^2+G_x^2+G_y^2}\begin{pmatrix} G_z\pm\epsilon && G_x-iG_y \end{pmatrix}
    \end{split}
\end{equation}
which satisfy the normalization conditions $\langle u_\pm^R|u_\pm^R\rangle=\langle u_\pm^L|u_\pm^R\rangle=1$.

\section{Additional Data From Simulation}
In this section, we provide additional analysis on the deviation between the analytical and numerical results \cite{hudata2025} presented in Fig. 2(c,d) of the main text. The previous theory in Ref. \cite{xu2017} includes up to the first order in the external force, while our theory, which arises from the first-order perturbation theory includes up to the 2nd order in force. Therefore, we expect that the deviation from the first-order theory has a leading order of $F^2$, while the deviation from the second-order theory has a leading order of $F^3$.

\begin{figure}[t]
    \centering
    \includegraphics[width=0.9\textwidth]{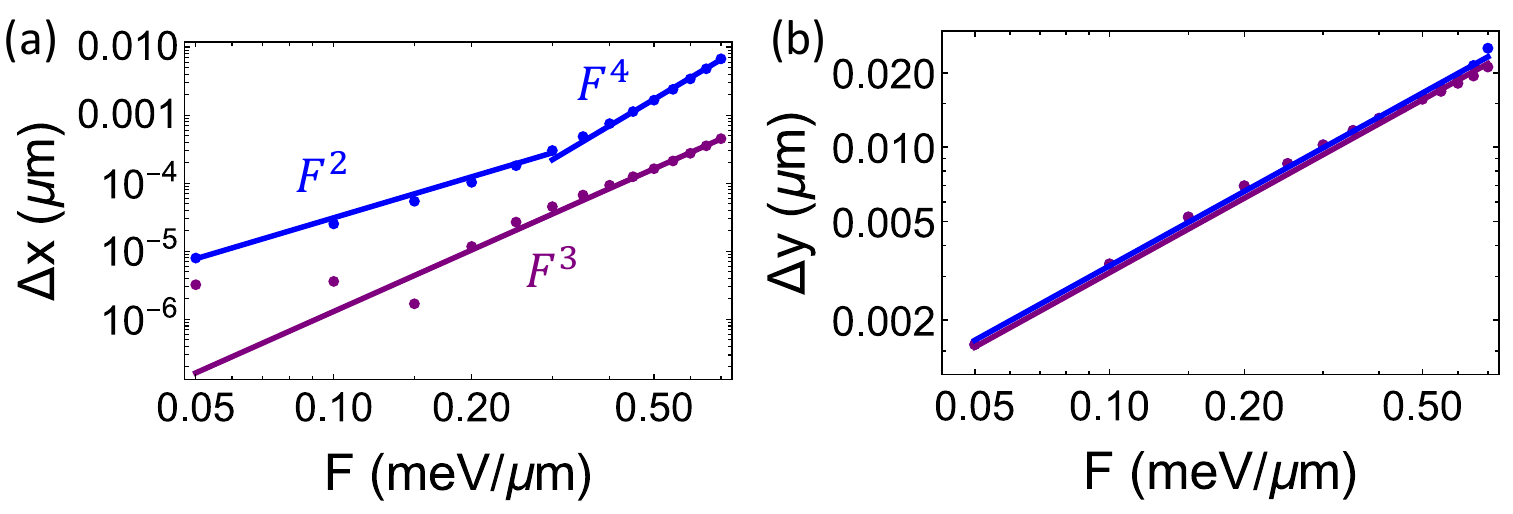}
    \caption{The deviation in position along x (a) and along y (b) between the numerical simulation  and the first order theory (blue dots), and the second order theory (purple dots), respectively, as a function of increasing external force $F$ at constant $t=1.25$ ps . The deviation from the first-order theory in $x$ was fitted with a quadratic function at lower $F$ and with a quartic function at higher $F$, while the deviation from the second-order theory (this work) is fitted with the cubic function in $F$. The deviations in $y$ in (b) from both theories are linear in $F$ (blue and purple lines, respectively).} 
    \label{fig}
\end{figure}

In Fig. \ref{fig}, we present the deviation in the wave-packet center-of-mass positions at $t=1.25$ ps plotted against the external force in log-log scale. At lower $F$, the deviation from the first-order theory is quadratic. However, the quadratic function fails to capture the data at higher $F$ and instead, a quartic function is required. This suggests that the higher-order terms becomes stronger at high $F$, and start to dominate over the deviation between the first-order theory and the simulation. On the other hand, the deviation from the second-order theory in $x$ does indeed exhibit cubic behaviour.

The first-order and the second-order theory gives almost the same result for the deviation in $y$, which can be fitted with a linear function in $F$. We believe that this is because the main correction comes from the correction to the anomalous Hall drift in the $x$-direction since the force is in the $y$-direction. It would require a new theory to capture the correction to the non-Hermitian anomalous Berry connection, which is beyond the scope of this work.

